# ADMM Based Semi-Structured Pattern Pruning Framework for Transformer

Tianchen Wang
Fudan University
Shanghai, China
21307130031@fudan.edu.cn

*Abstract:* **NLP (natural language processing) has achieved great success through the transformer model. However, the model has hundreds of millions or billions of parameters, which is a huge burden when deployed on personal computers or small-scale servers. To deal with it, we either make the model's weight matrix relatively sparser or compress the attention layer. Pattern pruning, one of the most important pruning methods, permits selecting and pruning a fixed number of parameters in each divided pattern block. However, the effect of pattern pruning is strictly limited by the sparsity within a region of weights in each layer. In this paper, we first introduced the Alternating Direction Method of Multipliers (ADMM) based pattern pruning framework to reshape the distribution of the activation map. Specifically, we propose to formulate the pattern pruning on the transformer as a constrained optimization and use ADMM to optimize the problem. In this way, the initial dense feature maps are transformed into rather regionally sparsified ones. Therefore, we can then achieve a higher compression ratio with better performance based on the pattern pruning method. Additionally, this paper provides theoretical derivations of the ADMM with local sparsity. Finally, we also extend the proposed ADMM-based framework with Sparse-Refined Straight-Through Estimator (SR-STE) to demonstrate its generalization and to avoid the gradient vanishing problem. We conduct extensive experiments on classification tasks over GLUE datasets. Significantly, we achieve a 50% compression ratio while maintaining an overall score of 80.1 on the GLUE dataset.**

## I. INTRODUCTION

The large model, especially the language large model (LLM), has achieved great success for its ability to learn from large amounts of data and efficiently train on large-scale GPU clusters. Its outstanding performance is due to the proposition of Transformers [1]. A transformer is a kind of structure that only uses multi-head attention to process the patched data and thus greatly improves the throughput volume. LLM can finish a series of downstream tasks with a relatively universal pre-trained model and simple fine-tuning. Therefore, LLM is expected to be applied in different devices and scenarios. However, it faces critical problems because of its complex model structure and great scale of parameters.

To ensure performance, LLM models usually maintain a great number of parameters. Even relatively small models [2, 3] consist of hundreds of millions of parameters, while bigger ones boost to billions. These LLMs are hard to deploy on the embedding device or mobile device. These challenges motivate researchers to find more effective techniques to compress [4] and design accelerators, like AWQ [5], to accelerate these models for these memory-bounded devices. These techniques include model quantization [6], low-rank decomposition [7], knowledge distillation [8], model sparsification and many others. They're often combined to compress model parameters [9]. This paper designs a better framework from the perspective of pruning to achieve improved compression effects.

Pruning is one of the great solutions of model sparsification, which can be categorized into unstructured pruning [10-12], structured pruning [13-15] and semi-structured pruning [16-19]. The basic theory of pruning is to remove the parameter, which is around zero, meaning it is of low importance. Unstructured pruning directly removes these individual parameters, which usually maintains the best accuracy but is unfriendly to the hardware implementation because of its irregularity. Structured pruning [13] removes parameters in a coarse way, which significantly eases the hardware burden but may reduce the accuracy. Semi-structured pruning maintains the accuracy and the sparsity of specified granularity, like block pruning [34]. This pruning method has been widely used in the accelerating of deep learning networks, like R-TOSS [35], an object detection framework. Its prominent effect has been widely realized and proved [36].

We take our eyes on the pattern pruning method [17], a semi-structured pruning method, and improve it to accelerate the Transformer model. It cuts the weight matrix into several blocks. In each block, the parameters will be first removed through unstructured pruning (e.g., a 16×16 matrix can divided into four 4*4 size ones). After clustering these sparse matrices and generating the pattern pool, the original dense matrix will search the best-matched sparse matrix in the pool and dot product with it to get the final pruned matrix. The whole process is shown in Fig.1. This pruning method simultaneously achieves high sparseness, hardware friendliness and good accuracy. Remarkably, pattern pruning can degenerate to the N×M pruning [16,18,20], whose sparse schemes can be searched stably [37]. However, directly applying pattern pruning on Transformers meets many problems, including low training speed and accuracy decline, with high sparseness due to its bigger weight and attention matrix size than CNN. Therefore, we think it's worth more optimization.

We gain the motivation to choose ADMM to optimize pattern pruning for the weight matrix of Bert (and other transformer models) with poor pattern sparsity, which refers to the submatrixes divided by pattern pruning as unbalanced for distribution on the aspect of sparsity. Thus, the effect of

ADMM can alleviate the phenomenon and change the polarized distribution in this pre-pruning step into the shape that frankly spreads around 50% of sparsity while maintaining accuracy. That's the core of the ADMM application, which boosts the efficiency of pruning.

In this paper, we present a systematic ADMM [21] based pattern pruning Transformer framework. We make the following contributions: (1) To our knowledge, it's the first time applying the pattern pruning on both the FFN and Attention layer to achieve a bigger compression ratio. We first introduce a Sparse-Refined Straight-Through Estimator (SR-STE [20]) in the Transformer pattern pruning training process to deal with the vanishing gradient and accuracy degradation problem. (2) We first respectively formulate the pattern pruning problem as a constrained nonconvex problem. We adopt the Alternating Direction Method of Multipliers (ADMM) [21] to systematically solve pattern pruning optimization problems on Transformer. By using ADMM, the optimization problem is then decomposed into two smaller sub-problems, which can be solved iteratively. In the pattern pruning problem, one of their sub-problems can be solved by stochastic gradient descent, and the other can be solved through a closed-form solution derived from the equation. (3) With extensive experiments on the GLUE dataset, we make a detailed analysis of the effect of our framework powered by ADMM and SR-STE are efficient in optimizing pattern pruning on the whole Transformer model. These optimization methods complement other compression methods, including knowledge distillation [8].

## II. RELATED WORKS

Due to the large scale of Transformer models, there is a growing interest in compressing these models to reduce their parameter size. Some studies [22] have shown that the Transformer model has found many redundant heads with embedded weights, which can be effectively pruned with minimal impact on accuracy. Pruning [17] represents a prevalent approach among the various strategies being explored for model compression. Non-structure pruning was proposed by Song Han [4], energy-aware pruning [23], and a series of pruning methods were proposed. Pattern pruning is a semi-structured pruning method that has also been used in DNN [17]. PatDNN first introduces the complete pattern pruning method (including different kinds of optimization) and applies it to DNN workloads. Its theory is akin to N:M Transformer [16]: divide the weight matrix into sub-matrices of a specified size. Within each sub-matrix, unstructured pruning of fixed sparsity is then applied. However, N:M pruning possesses its own unique features: N:M pruning specifically targets the removal of M contiguous parameters within a sequence of N parameters, typically located in a single row or column. In contrast, the divisions of common pattern pruning are more flexible, which divides the weight matrix into the given size visually (e.g., 3*3). The N:M Transformer made an attempt to apply the ADMM to the NLP (BERT) model and showed a good effect on the pruning of ADMM. From the performance and resources respect, we learn from N:M Transformer that ADMM can be widely applied to other pruning method on NLP and Transformer model.

The PP-Transformer [19] innovatively incorporates pattern pruning into the Transformer Model and implements it on FPGA hardware. However, it currently focuses only on applying pattern pruning to the dense weights of MLP layers and lacks optimization, leaving room for further improvement. Our work extends the PP-Transformer and PatDnn methods by utilizing ADMM and SR-STE compression techniques. We specifically optimize pattern pruning and apply it to the attention layer at the algorithm level, improving the efficiency of the Transformer model.

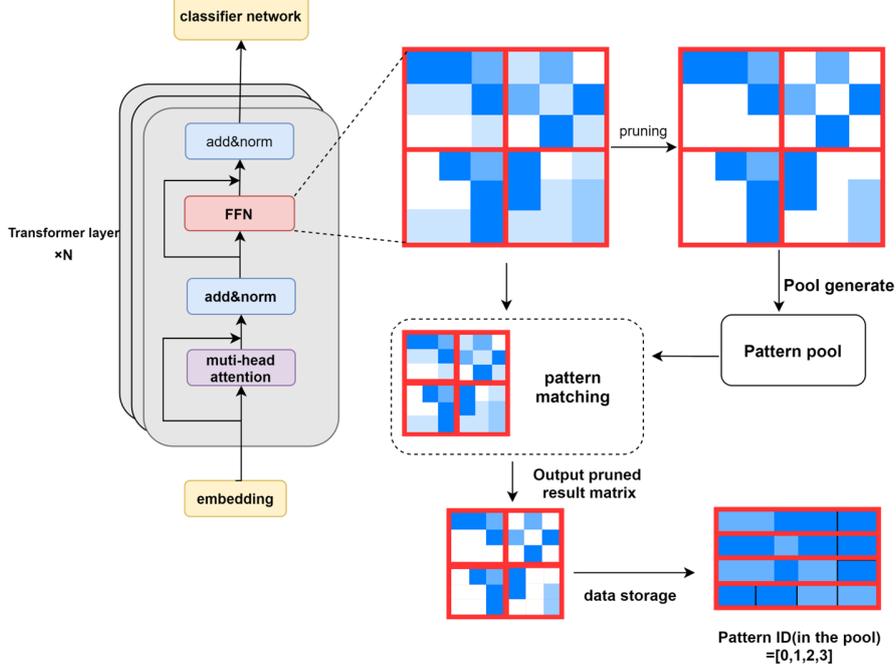

Fig. 1. An illustration of pattern pruning applied in the Transformer model

Unlike the N:M Transformer that leverages sparse N×M NVIDIA tensor cores for accelerated computations across different dimensions, our framework utilizes ADMM and other techniques to achieve competitive performance. We have successfully developed a custom FPGA hardware accelerator specifically designed to optimize pattern pruning and other techniques. This accelerator greatly improves the hardware-friendliness of the entire transformer block, including the attention mechanism. Furthermore, models incorporating our framework, such as PP-Transformer, can be utilized with TensorRT [24]. TensorRT includes optimized operators that effectively leverage hardware features, maximizing performance gains. Similar tools like Flopoco [25] can efficiently generate the entire Transformer model and the corresponding RTL code of operators. The availability of such tools demonstrates that the hardware implementation of our ADMM-based pattern pruning Transformer framework is both feasible and relatively straightforward.

## III. METHODOLOGY

In this section, we first provide an introduction to the background of ADMM for neural networks, establishing the foundation for our approach. Then, we outline how we formulate the problem, clearly defining the objectives and constraints of our study. Next, this paper presents an overall methodology, explaining how we leverage ADMM to develop a comprehensive framework for pattern pruning in Transformer models. Furthermore, we delve into the specific application of pattern pruning to the attention mechanism. Finally, we provide a detailed overview of the training process, describing the steps involved in training the pattern-pruned Transformer model using our proposed framework.

### A. Background of ADMM for Neural Networks

The Alternating Direction Method of Multipliers (ADMM) is a universal optimization method. It is primarily to solve the minimization of the objective function with multiple variables, which have constraints among each other. Its biggest advantage is to fully use the objective function's decomposability to iteratively optimize several variables. ADMM simultaneously benefits from dual ascent and multiplier methods. When facing a large-scale or specific problem, it can decompose the overall problem into smaller subproblems, solve them in parallel, and aggregate them to get the complete solution. Now, it has been widely used in neural networks [17,26,27,30], primarily to help compress convolution neural networks because of its powerful ability to cope with the optimization of loss function. A imaging framework [31] that blends the advantages of ADMM and deep learning also has been proposed and inspires us.

### B. Problem definition

We consider a Bert model with 12 transformer layers, which include muti-head attention and Feed Forward Network (FFN). Our target is to compress all FC (fully-connected) layers. Collect the weight matrix in FFN as $W_i$ and the bias of weight as $b_i$, where i is the layer number. The loss function of the Bert model is denoted by $f(\{W_i\}_{i=1}^N, \{b_i\}_{i=1}^N)$. N represents the total number of layers.

Universally, we use ADMM to solve this optimization problem (on most Neural Networks) like:

$$\text{minimize} \quad f(\{\mathbf{W}_i\}_{i=1}^N, \{\mathbf{b}_i\}_{i=1}^N), \quad \text{subject to} \quad \mathbf{W}_i \in \mathbf{S}_i, i = 1, \dots, N \quad (1)$$

Due to the diversity of the choice of $S_t$, ADMM can be applied to the pattern pruning. For the pattern pruning, the choice of $S_i$ is a set that contains several small matrices derived from the decomposition of larger matrices. Each small matrix in the set has a fixed sparsity. The sparsity can be manually set and changed.

### C. Framework based on ADMM

The above problem is nonconvex, and it is impossible to optimize it only with the SGD (stochastic gradient descent). However, by using the framework based on ADMM, this problem can standardly be solved using an alternating optimization approach. In this way, we have to first introduce a specific indicator function:

$$g_i(\mathbf{W}_i) = \begin{cases} 0 & \text{if } W_i \in \mathbf{S}_i \\ \infty & \text{otherwise} \end{cases} \quad (2)$$

We then introduce the auxiliary variable $\mathbf{Z}_i$ and rewrite the problem as

$$\underset{\{\mathbf{W}_i\},\{\mathbf{b}_i\}}{\text{minimize}} \quad f(\{\mathbf{W}_i\}_{i=1}^N, \{\mathbf{b}_i\}_{i=1}^N) + \sum_{i=1}^N g_i(\mathbf{Z}_i), \quad \text{subject to} \quad \mathbf{W}_i = \mathbf{Z}_i, i = 1, \dots, N \quad (3)$$

Then, through the application of the augmented Lagrangian method, the whole problem can be decomposed into two sub-problems. The first problem is:

$$\text{minimize}_{\{\mathbf{W}_i\},\{\mathbf{b}_i\}} \quad f(\{\mathbf{W}_i\}_{i=1}^N, \{\mathbf{b}_i\}_{i=1}^N) + \sum_{i=1}^N \frac{\rho_i}{2} \| \mathbf{W}_i - \mathbf{Z}_i^k + \mathbf{U}_i^k \|_F^2 \quad (4)$$

The $\rho_i$ is the hyper-parameter. The $\mathbf{U}_i^k$ is the dual variable updated in every iteration, and the k in $\mathbf{U}_i^k$ presents the $k_{th}$ iteration. Its updated formulation is: $\mathbf{U}_i^k := \mathbf{U}_i^{k-1} + \mathbf{W}_i^k - \mathbf{Z}_i^k$. This problem becomes convex and can be solved by stochastic gradient descent. We only need to add the extra loss $\sum_{i=1}^N \frac{\rho_i}{2} \| \mathbf{W}_i^{k+1} - \mathbf{Z}_i + \mathbf{U}_i^k \|_F^2$ generated by ADMM, and train the model normally.

The second problem is:

$$\underset{\{\mathbf{Z}_i\}}{\text{minimize}} \sum_{i=1}^N g_i(\mathbf{Z}_i) + \sum_{i=1}^N \frac{\rho_i}{2} \| \mathbf{W}_i^{k+1} - \mathbf{Z}_i + \mathbf{U}_i^k \|_F^2 \quad (5)$$

Though this problem is convex, it can be smartly dealt with by applying $\mathbf{Z}_i$ with Euclidean project onto the set $S_i$.

$$\mathbf{Z}_i^{k+1} = \Pi_{\mathbf{S}_i}(\mathbf{W}_i^{k+1} + \mathbf{U}_i^k) \quad (6)$$

It can trivially be seen that it makes $g_i(\mathbf{Z}_i)$ equal to zero while maintaining the distance between $\mathbf{Z}_i$ and $\mathbf{W}_i^{k+1} + \mathbf{U}_i^k$ as short as possible.

Based on the different problems, the Euclidean projection is variant. For the single pattern pruning problem, $Z_i^{k+1}$ equals to the result matrix of $\mathbf{W}_i^k + \mathbf{U}_i^k$ after pattern pruning. That is to say, the matrices obtained by dividing a matrix exhibit the same sparsity. For the single quantization problem, $Z_i^k + 1$ equals the result of setting all the values in $\mathbf{W}_i^{k+1} + \mathbf{U}_i^k$ as quantization values. For the ADMM optimization on pattern pruning, we do several experiments and compare the results of a direct implementation with the results obtained after employing the ADMM framework.

*D. Pattern pruning of attention*

We not only choose the dense weight matrix of the model, but also the attention mechanism to apply the pattern pruning to extend our framework to with more sparsification. There are already many attempts to prune the attention, like [32,33]. In the attention layer, after the matrix multiplication of Query and Key, we do the pattern pruning to sparsify their results. By partitioning the attention map into multiple blocks, we can easily observe the relative importance of each block based on their numerical values. As shown in Fig. 2, the dense attention map is transformed into a sparse attention map through pattern pruning. The first and last blocks exhibit greater importance.

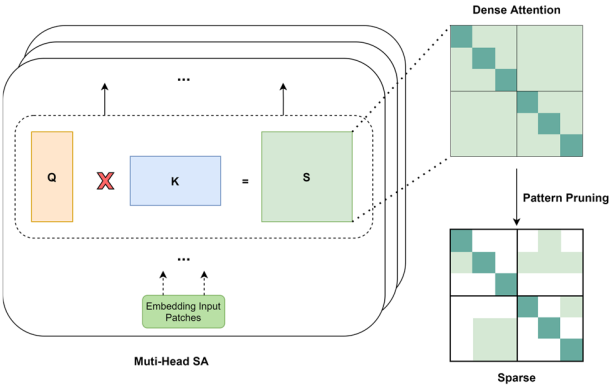

Fig. 2. Pruning process of attention

*E. The whole training process and SR-STE*

The whole training process is shown in Fig. 3. Firstly, we initialize both $Z_i^0$ and $U_i^0$ as zero matrices. In the first iteration, we gain the weight $W_i^0$. Then, we project $Z_i^0$ and update $U_i^0$. Next, we add the extra part generated by ADMM to the loss to deal with the first subproblem. We normally train the model and enter the next iteration. We repeat the above steps until the model converges. After ADMM optimization is done, we will keep the existing sparsity distribution. Then, we prune and retrain the model to fit the downstream task

SR-STE(Sparse-Refined Straight-Through Estimator) is introduced in the whole training process. It focus on the back propagation of the weight matrix. After a dense weight matrix $W_t$ has been pruned into a sparse weight $\widetilde{W}_t$, we will add the extra value in the loss, like the equation (7) shows.

$$W_{t+1} = W_t - \gamma_t(g(\widetilde{W}_t) + \lambda_W(\bar{\mathcal{E}}_t \odot W_t)) \qquad (7)$$

Here $W_{t+1}$ is the weight matrix after the update of original weight $W_t$. $\bar{\mathcal{E}}_t$ refers to the mask for the pruned weight, $\odot$ denotes Hadamard product, $\lambda_W$ refers to the relative weight for the sparse-refined term, $\gamma_t$ denotes the learning rate and $g(\widetilde{W}) = \nabla_{\widetilde{W}} \mathcal{L}(\widetilde{W})$, which is a gradient function based on the sparse network $\widetilde{W}$. The introduce of $\widetilde{W}_t$ into the loss greatly enhances the robustness and positively avoids the gradient vanishing problem.

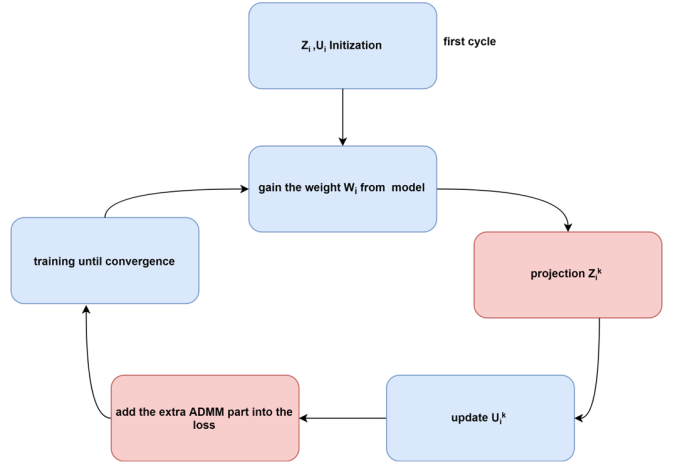

Fig. 3. The training process of a single optimization problem

## IV. EVALUATION

In this section, we evaluate the ADMM-based framework and show its effectiveness in compressing Transformer networks over a wide range of NLP tasks.

**Implementation**. We conduct experiments on a NVIDIA 3090 GPU with batch size 24 and use an Adamw optimizer [28] with an initial learning rate of 7e-5 while using a learning rate scheduler to make adaptive adjustments during the training process. The weight decay is set to 0.001. Both pre-training and pattern pruning were trained for 5 epochs. The experimental section uses Bert as the base model for pruning. In order to test the performance on different downstream tasks, we use Bert-base as an example. The data processing and other experimental settings remain consistent with Bert's default settings. The experimental environment consists of Python 3.9.3, PyTorch 1.13.1, and CUDA 12.3.

Former work PP-Transformer [19] has indicated that higher sparsity leads to greater performance loss when the pattern size is larger than 4\*4. Therefore, for evaluation, we focus on evaluating 50% sparsity, 4\*4 pattern size, 32 pattern number, as these settings are applicable in most situations. We use pre-trained model checkpoints for BERT2, provided by the HuggingFace model repository. All models require fine-tuning, and the time required for fine-tuning depends on the size of the dataset and tasks, with a maximum duration of 7 hours. We further choose 4\*4 pattern sizes with 50% sparsity to demonstrate the change in the weight distribution of the model.

**Datasets.** We evaluate the ADMM-based pattern pruning framework and our baselines using the General Language Understanding Evaluation (GLUE) benchmark [29], a collection of NLP tasks varying in data availability and complexity. We report the Spearman correlation for Semantic Textual Similarity Benchmark (STS-B), the F1 score for Microsoft Research Paraphrase Corpus (MRPC), Matthews correlation for the Corpus of Linguistic Acceptability (CoLA), and accuracy for all remaining tasks: MultiNLI (MNLI), the Stanford Sentiment Treebank (SST-2), Question NLI (QNLI), Recognizing Textual Entailment (RTE). The reported average is the geometric mean of reported scores. With GLUE as an evaluation criterion, we can better compare our framework with other pruning methods.

## A. Main results

We use Bert as the baseline model. To demonstrate the effectiveness of Pattern pruning (optimized with ADMM), we include ADMM unstructured, N× M Transformer, and Pattern pruning (without ADMM) as comparisons. Here is a detailed explanation of each method:

**Bert:** Baseline Bert base model

**ADMM (unstructured):** To measure the accuracy of the semi-structured method specifically, we utilize ADMM but introduce unstructured sparsity at a per-layer level instead of applying global sparsity. The objective is to achieve a 50% sparsity rate within each layer.

**N×M Transformer:** The transformer framework was pruned with the N:M method and optimized with ADMM.

**Pattern pruning (without ADMM):** The transformer framework is pruned using a pattern pruning method with no ADMM optimization.

**Pattern pruning (optimized with ADMM):** The transformer framework was pruned using a pattern pruning method optimized with ADMM.

In the experiment, we iteratively adjusted the value of $\rho$ and ultimately determined that setting $\rho$ to 0.01 yielded the best results across all tasks. Table 1 presents the results of different methods of BERT pruning, and we made the following observations: First, the ADMM (unstructured), a basic pruning-based method, sparsifies the weights of Transformer blocks but cannot explicitly satisfy the underlying hardware constraints, e.g., the pattern sparsity. While the highest accuracy on downstream tasks was preserved (81.3% on average compared to the baseline's 81.8%), the resulting sparse weights exhibited a random structure of non-zero weights. This random structure was inefficient and considered unacceptable. Second, the pattern pruning method (without ADMM optimization) demonstrated relatively poor performance, both in individual downstream tasks and on average optimization) demonstrated relatively poor performance (78.1%). This indicates that the method's effectiveness was limited without utilizing the ADMM technique.

In contrast, our method, pattern pruning optimized with ADMM, achieved superior results across various tasks, even approaching the performance of the N×M Transformer. It is worth noting that the N×M Transformer has undergone extensive optimization. Furthermore, our method is more flexible as it allows for setting the sparsity level for each layer, and even the distribution of each individual weight matrix. Notably, it achieved the best scores of 55.4% and 68.8% on the COLA and RTE tasks, surpassing the N×M Transformer. The overall average score reached 80.1%, which was significantly higher than that of the pre-optimization pattern pruning method by nearly 2%.

Table 1. Results on GLUE benchmark

| Model | Task | | | | | | | Average |
|---|---|---|---|---|---|---|---|---|
| | MNLI (m/mm) 392k | SST-2 67k | QNLI 108k | CoLA 8.5k | STS-B 5.7k | MRPC 5.5k | RTE 2.5k | |
| Baseline | 84.5/84.8 | 92.5 | 91.6 | 56.7 | 89.6 | 91.7 | 70.7 | 81.8 |
| ADMM (unstructured) | 84.0/84.7 | 92.5 | 91.0 | 57.5 | 89.6 | 90.5 | 68.2 | 81.3 |
| N×M Transformer | 82.3/83.4 | 92.3 | 90.4 | 55.3 | 89.3 | 90.8 | 68.6 | 80.5 |
| pattern pruning (without ADMM) | 80.6/81.5 | 90.9 | 88.8 | 53.7 | 87.3 | 85.1 | 65.3 | 78.1 |
| pattern pruning (optimized with ADMM) | 82.0/83.1 | 92.0 | 89.9 | **55.4** | 88.4 | 89.2 | **68.8** | 80.1 |

## B. Effect on weight distribution

Furthermore, we investigate the impact of our ADMM-based framework on the weight distribution of each layer. Fig. 4 illustrates the distribution changes of weights in the 0th layer of BERT when before/after using the ADMM-based framework. Our framework can effectively tailor to different shapes' weights and adhere to the division to purposefully make sparsity of each part approach our given sparsity. Here, we set it as 50%. In Fig. 4, the ADMM-optimized weight matrix (the query weight of layer 0) exhibits a notable increase in zero values compared to the original matrix. This demonstrates the effectiveness of the ADMM optimization in achieving a 50% sparsity distribution for pattern pruning.

What's more, the extensive Query weight matrix (768,768) is partitioned into nine discrete sections, each with a dimension of 256. This approach is employed to illustrate the impact of the ADMM optimization framework on the distribution of weights within each section. The weight distributions of the nine independent parts are presented in Fig. 5. It can be observed that the majority of these parts exhibit a similar trend to the original matrix, approaching 50% sparsity and becoming increasingly sparse. These findings prove the effectiveness of the proposed approach in altering weight distributions and enhancing hardware compatibility.

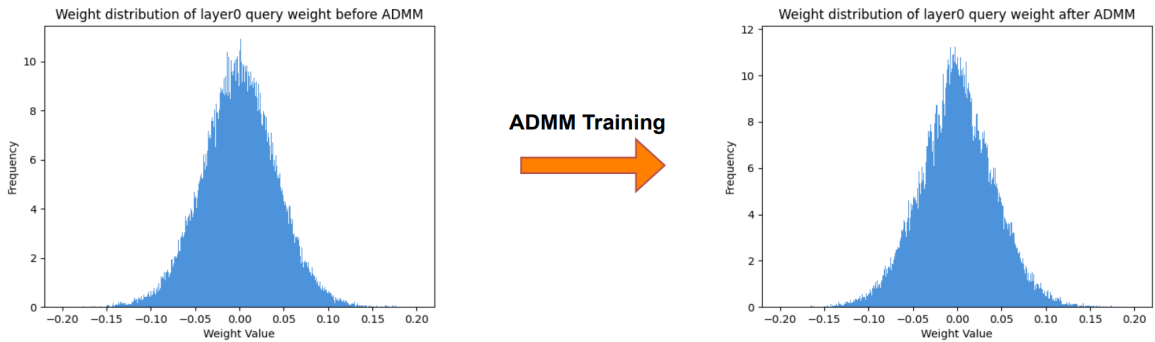

Fig. 4. Distribution change of the whole layer 0 query weight before/after ADMM for pattern pruning

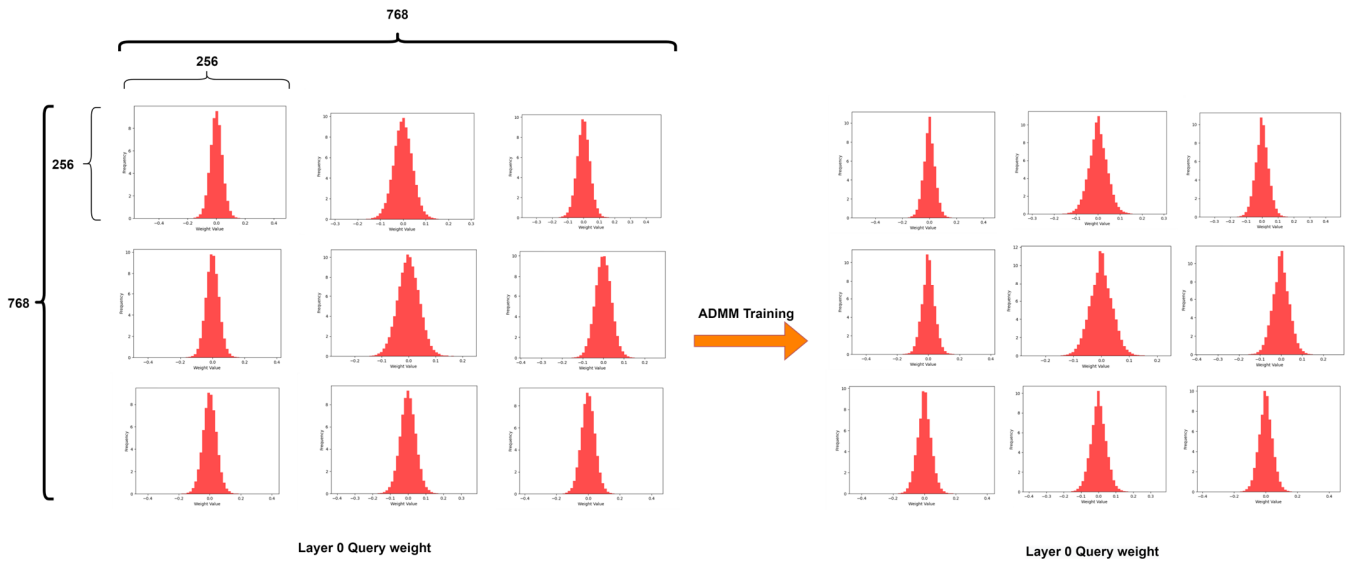

Fig. 5. Distribution of each part of the layer 0 query weight

*C. Analysis of results*

Our best accuracy of the Bert model after optimization, before pruning, achieves 81.8% on average. After fully pruning the model, the average score of the Bert model can achieve 80.1% while maintaining 50% of sparsity under ADMM optimization. For specific tasks, COLA and RTE can outperform the N×M Transformer, which is an excellent framework.

**Validation and hardware-friendliness improvement.** Our evaluation scores and hardware efficiency are clearly demonstrated by quantitative experiments. The significant improvements can be attributed to our ADMM optimization framework. Within this framework, we integrated the SR-STE method to prevent gradient vanishing, enhancing the training effect. During pattern pruning, varying sparsity levels in different weight blocks led to substantial discrepancies between the pruned matrices and the original large matrices, severely affecting validation accuracy. Our ADMM-based framework iteratively optimizes the weight distribution, ensuring that the weights are more evenly distributed to approximately 50% sparsity before pattern pruning. This approach significantly boosts the post-pruning scores, achieving an overall score of 80.1, which is 2 percentage points higher than pre-optimization and only 0.4 percentage points lower than one of the best current sparse frameworks, the N×M Transformer. Additionally, the structural uniformity achieved through our pruning method provides hardware efficiency comparable to the N×M Transformer. By leveraging and improving the existing pattern-pruning FPGA hardware framework, we can achieve excellent deployment results.

**Scalability of the framework and comparsion with state-of-the-art model compression techniques.** When applying our ADMM-based pattern pruning framework on larger models, like GPT, it is estimated to be less efficient for its restriction of pattern pruning itself, whose pruning process is too long for those large weight matrix due to the pattern clustering and the mask applying on the weight matrix. The applying of our framework on edge devices or real-time applications is prospective because our framework can quickly fit for these devices by pruning the Transformer model in specific sparsity and changing the weight distribution into the best shape for its deployment. Therefore, our framework remains potential impact in different deployment scenarios. Meanwhile, although the state-of-the-art model compression techniques, including

Mixture of Experts(MoE) [39], knowledge distillation [8] are popular, the traditional pruning method still retain its postion of one of the most important compression technique for its low universality for different kinds of models and low requirement for computing power. Our ADMM-based pruning framework will certainly hold its importance on its more universal applying on all Transformer models to compress them and get a smaller while more effective and sparse pre-train model.

## V. FUTURE WORKS

According to PP-Transformer [19], the performance of pattern pruning with the bigger division (e.g. 8×8) will drop quickly when sparsity becomes above 80%. In our experiment, we observed that even after applying ADMM improvement, the performance degradation with high sparsity and larger pattern choice pattern pruning remained minimal but still exhibited an unacceptable drop in performance. Therefore, we want to find another optimization method and combine it with ADMM to completely stop it from dropping. On the other hand, we hope we can develop a complete set of hardware frameworks tailored to ADMM optimization, which will greatly speed up the deployment and run time on FPGA or ASIC chips. We also hope to find a automatic and general method to quantize the detailed sparsity extent and the dependency between layers and cutted group coupled parameters, like Depgraph [38], and implement this method for semi-structured pruning.

## VI. CONCLUSIONS

ADMM can reshape the weight distribution of the transformer to the desired form and significantly enhance the pruning performance. This paper demonstrates the effectiveness of ADMM on pattern pruning from performance and weight distribution aspects. Furthermore, the ADMM-based pattern pruning framework is orthogonal to many other optimization techniques. This framework we proposed can be widely applied to almost all the models based on the transformer, including Bert, ViT and others. While the framework introduces an accuracy gap due to compression, we firmly believe that the ADMM-based pattern pruning framework will continue to be effective and maintain its significance.

## ACKNOWLEDGEMENT

I sincerely thank my advisor ZH.L and the anonymous reviewers for their constructive comments.